\documentclass{aa}  

\usepackage{graphicx}
\usepackage{txfonts}
\usepackage[]{hyperref}
\usepackage{epstopdf}
\usepackage{todonotes}
\usepackage{multirow}
\newdimen\mylength
\begin{document}

   \title{Weak lensing analysis of galaxy pairs using CS82 data}

\author{Elizabeth Johana Gonzalez \thanks{ejgonzalez@unc.edu.ar }\inst{1,2} \and Facundo Rodriguez \inst{1,2} \and Diego Garc\'ia Lambas \inst{1,2} \and Mart\'in Makler \inst{3} \and
Valeria Mesa\inst{4} \and Sol Alonso \inst{5} \and Fernanda Duplancic \inst{5} \and Maria E. S. Pereira \inst{6} \and  HuanYuan Shan\inst{7}}

\institute{Instituto de Astronom\'{\i}a Te\'orica y Experimental, (IATE-CONICET),
Laprida 854, X5000BGR, C\'ordoba, Argentina
        \and
  Observatorio Astron\'omico de C\'ordoba, Universidad Nacional de C\'ordoba, Laprida 854, X5000BGR, C\'ordoba, Argentina
  \and
Centro Brasileiro de Pesquisas F\'{\i}sicas (CBPF), Rio de Janeiro, RJ 22290-180, Brasil 
  \and  
 Instituto Argentino de Nivolog\'ia, Glaciolog\'ia y Ciencias Ambientales (IANIGLA-CONICET), Parque Gral San Mart\'in, CC 330, CP 5500, Mendoza, Argentina
   \and
Departamento de Geof\'isica y Astronom\'ia, CONICET, FCEFyN, UNSJ, Av. Ignacio de la Roza 590 (O), J5402DCS, San Juan, Argentina
    \and
Brandeis University, 415 South Street, Waltham, MA 02453, USA
    \and
    Shanghai Astronomical Observatory (SHAO), Nandan Road 80, Shanghai 200030, China
             }

   \date{2017}

\abstract{In this work we analyze a sample of close galaxy pairs (relative projected separation $<25$\,$h^{-1}$\,kpc and relative radial velocities $<$ 350 \,km $s^{-1}$) using a weak lensing analysis based on the CFHT Stripe 82 Survey. We determine halo masses for the \textit{Total sample} of pairs as well as for \textit{Interacting}, \textit{Red} and \textit{Higher luminosity} pair subsamples with $\sim 3\sigma$ confidence. The derived lensing signal for the total sample can be fitted either by a singular isothermal sphere with $\sigma_V = 223 \pm 24$ km\,s$^{-1}$ or a NFW profile with $R_{200} = 0.30 \pm 0.03\,h^{-1}$\,Mpc. The pair total masses and total $r$ band  luminosities imply an average mass-to-light ratio of $\sim 200\,h\,M_\odot/L_\odot$.  On the other hand, \textit{Red} pairs which include a larger fraction of elliptical galaxies, show a larger mass-to-light ratio of $\sim 345\,h\,M_\odot/L_\odot$. Derived lensing masses were compared to a proxy of the dynamical mass, obtaining a good correlation. However, there is a large discrepancy between lensing masses and the dynamical mass estimates, which could be accounted by astrophysical processes such as dynamical friction, by the inclusion of unbound pairs, and by significant deviations of the density distribution from a SIS and NFW profiles in the inner regions. We also compared lensing masses with group mass estimates obtained from the \citeauthor{Yang2012} galaxy group catalog, finding a very good agreement with the sample of groups with 2 members. \textit{Red} and \textit{Blue} pairs show large differences between group and lensing masses, which is likely due to the single mass-to-light ratio adopted to compute the group masses.} 
 

   \maketitle
%

\section{INTRODUCTION}
Studies  of galaxy pairs are important in order to understand the formation of massive galaxies and groups. Galaxy interactions in these systems are very common and play a significant role in galaxy evolution \citep[e.g.][]{Woods07, Ellison2010,Mesa2014}, since they can produce galaxy morphology transformations \citep{Toomre72, Barnes92, Hopkins08}. For instance, the presence of a close galaxy companion drives a clear enhancement in galaxy morphological asymmetries, and this effect is statistically significant up to projected separations of at least $<50$\,$h^{-1}$\,kpc \citep{Patton2016}. Also, interacting galaxy pairs show physical effects generated by the interaction which can be detected at large projected separations, as far as $\sim 150$\,kpc \citep{Patton13}. These effects include enhancement of the star formation rate \citep{Ellison08, Ellison2010, Patton11, Scudder12,Lambas2012,Patton13} and AGN production \citep{Barnes91,Hopkins08,Darg10}, since it has been shown that the fraction of AGNs is larger for galaxies in pairs \citep{Alonso2007}. Despite the importance and prevalence of galaxy mergers in driving galaxy evolution, the physical details of the merging process are not yet fully understood even in the local Universe.

Mass determinations of halos hosting galaxy pairs can contribute to get a better understanding regarding galaxy evolution. A dynamical approach is commonly applied to estimate the mass enclosed by the galaxies. Nevertheless, determining the virial masses is very difficult for these systems given that the only available information is the velocity difference along the line-of-sight and the projected distance between the galaxies. Given the limitations, various methods have been proposed \citep{Nottale2018,Chengalur1996,Peterson1979,Faber1979}, however recovering the mass remains very
uncertain. 

Gravitational lensing allows us to derive the total projected mass distribution directly without relying on visible tracers. Strong lensing in particular provides information on the inner regions of galaxy systems \citep[e.g. ][]{Collett2017,Cerny2018,Jauzac2018} and galaxies \citep[e.g. ][]{Ven2010,Dutton2011,Suyu2012}. \citet{Shu2016} applied this technique to model the mass distribution of a galaxy pair lens system, obtaining significant spatial offsets between the mass and light of both lens galaxies, which could be related to the interactions between the two lens galaxies. On the other hand, weak gravitational lensing is a powerful statistical tool that provides information regarding the projected mass distribution of galaxy systems at larger distances.

In particular, stacking techniques allow one to study low mass galaxy groups and to derive the properties of the combined systems \citep[e.g. ][]{Leauthaud10,Melchior13,Rykoff08,Foex14, Chalela2017, Chalela2018}. This technique has also been successfully applied to measure the mean masses of dark matter halos for stacked samples of isolated galaxies \citep[e.g. ][]{Mandelbaum2006,Uitert2011,Schrabback2015,Charlton2017}. Therefore, weak gravitational lensing is useful to study the dark matter halos of a wide range of masses and can be applied to analyze the total mass content of galaxy pair systems. 

In this work we analyze a sample of close galaxy pairs with a relative projected separation, $r_p$, smaller than 25\,$h^{-1}$\,kpc and relative radial velocities $\Delta V <$ 350 \,km $s^{-1}$. This enhances the likelihood that the pair is physically associated as it reduces the number of pairs identified due to projection effects \citep{Alonso2004,Nikolic2004,Edwards2012}. Also \citet{KitzbichlerWhite2008} identified close pairs in mock galaxy catalogues within  $\Lambda$-CDM  cosmological simulations, finding that most  pairs identified with a projected separation $r_p\leq$ 25\,$h^{-1}$\,kpc and radial velocity difference $\Delta V <$ 300 \,km $s^{-1}$  are physically close and  will merge in an average time of $\sim$2 Gyrs.  Moreover, at larger separations ($r_p > 100$\,kpc) it is important to consider the competing influences of other neighbouring galaxies \citep{Moreno2013,Karman2015} and  larger scale environmental influences \citep{Park2007,Moreno2013,Sabater2013}. We derive total masses through weak lensing using the CFHT Stripe 82 Survey (CS82, Erben et al. in prep.) for different galaxy pair samples. Derived masses are compared to an indicative dynamical mass and with masses obtained from a galaxy group catalog computed according to the mass-to-light ratio. The work is organized as follows: In Sect. \ref{sec:data} we describe the sample of galaxy pairs. In Sect. \ref{sec:lensing} we give the details of the lensing analysis. In Sect. \ref{sec:results} we present the obtained masses, the mass-to-light ratios for the different galaxy pair samples analyzed and the comparison to other mass estimates. Finally in Sect. \ref{sec:conc} we summarize and discuss the results. We adopt when necessary a standard cosmological model with $H_{0}$\,=\,70\,km\,s$^{-1}$\,Mpc$^{-1}$, $ \Omega_{m} $\,=\,0.3, and $ \Omega_{\Lambda} $\,=\,0.7.

\section{GALAXY PAIR SAMPLE}
\label{sec:data}

\begin{figure}
\centering
\includegraphics[scale=0.45]{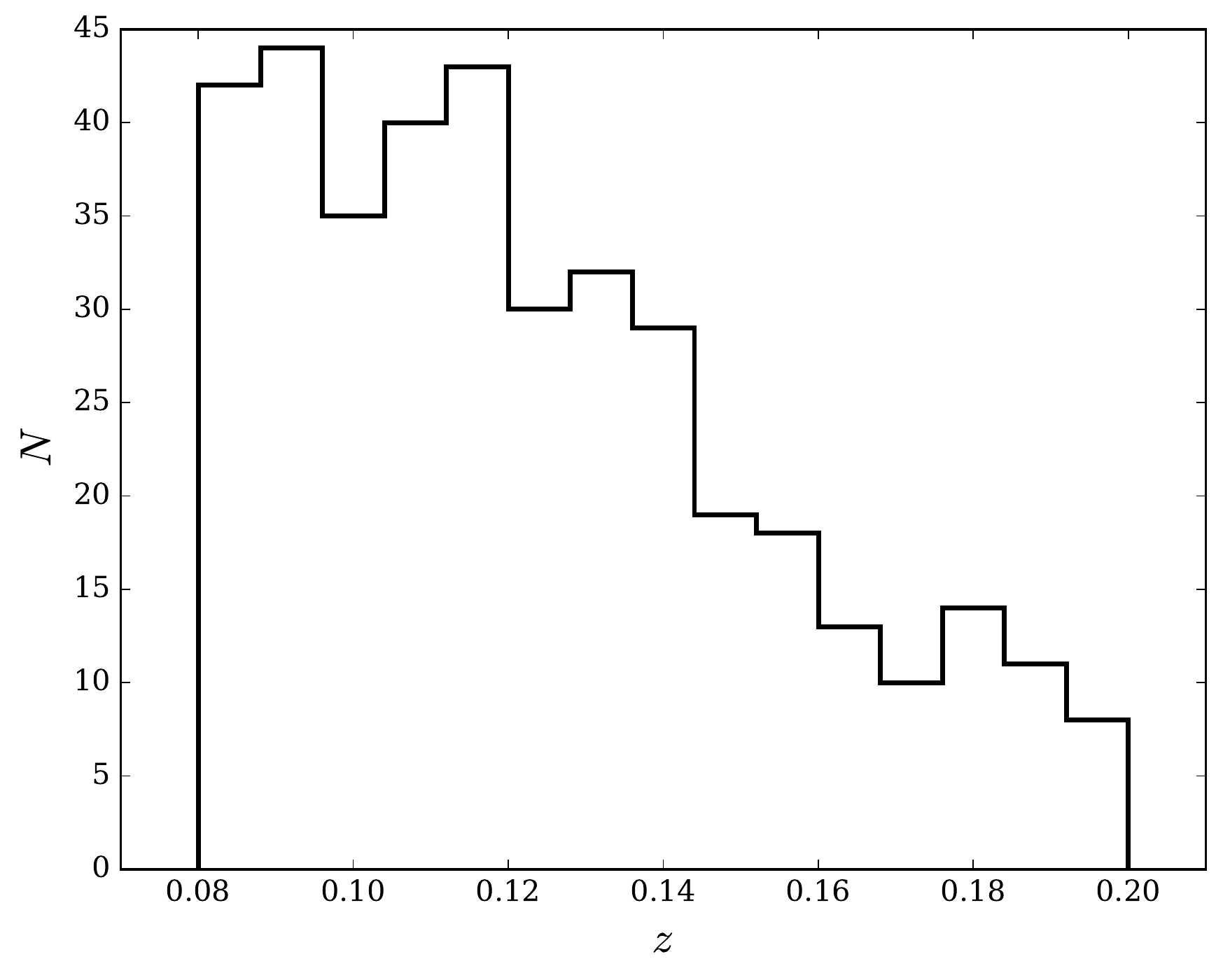}
\includegraphics[scale=0.51]{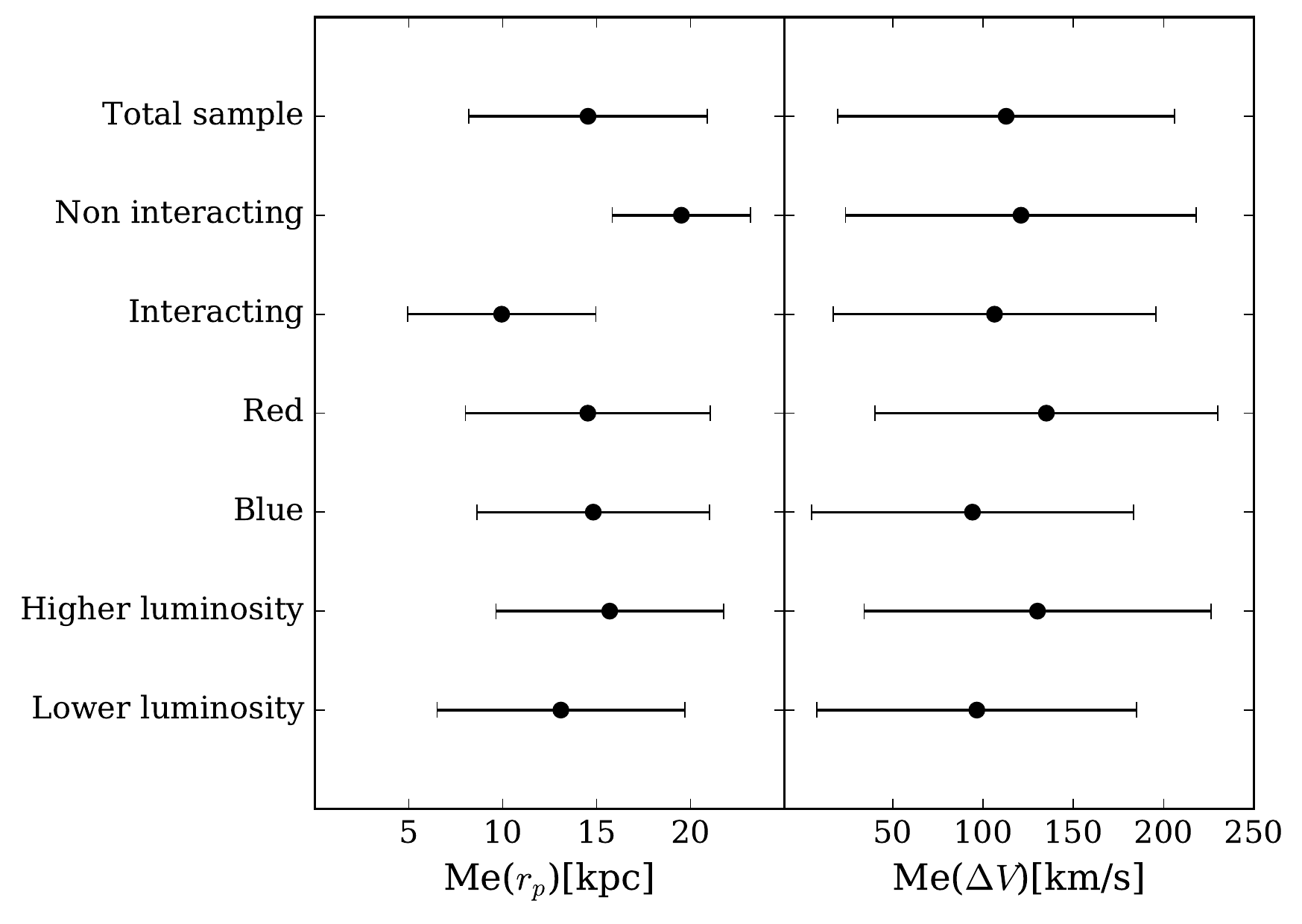}
\caption{\textit{Upper panel:} Redshift distribution of the total sample of galaxy pairs. \textit{Lower panel:} Median projected distances, Me$(r_p)$, and median radial velocity difference, Me($\Delta V$), for the different galaxy pair classes. Error bars corresponds to the standard deviations of $r_p$ and $\Delta V$ distributions obtained for each class. \textit{Non interacting} pairs tend to have larger $r_p$ distances than \textit{Interacting} pairs. On the other hand, $r_p$ distributions for \textit{Red} and \textit{Blue pairs}, as well as \textit{Higher} and \textit{Lower luminosity} pairs, are all in agreement with the total sample $r_p$ distribution.} 
\label{fig:pares_dist}
\end{figure}

\begin{figure}
\centering
\includegraphics[width=0.48\textwidth]{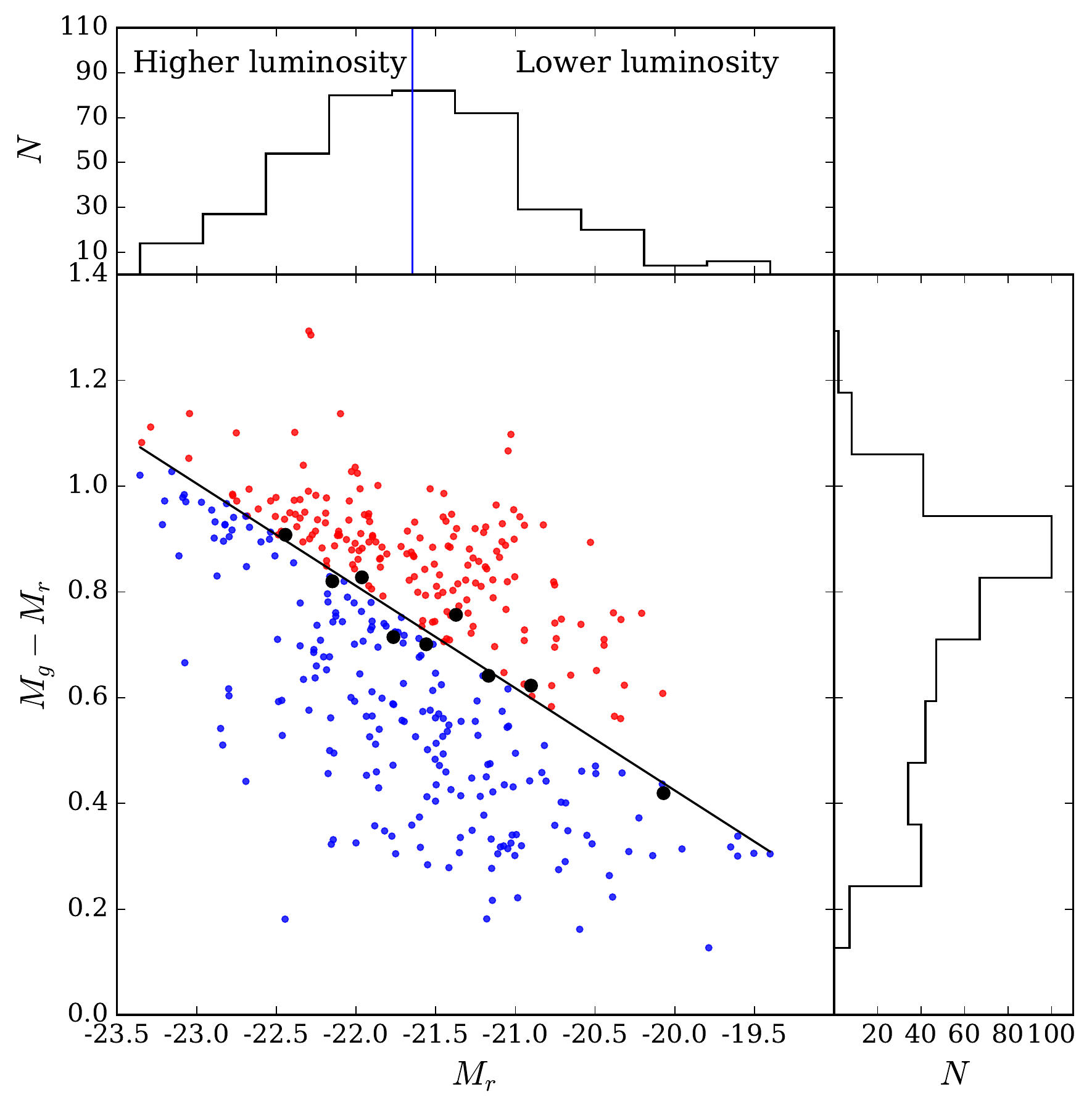}
\caption{Red and Blue pairs classification criterion (\textit{central panel}): $M_g-M_r$ vs. $M_r$, where $M_g$ and $M_r$ are the total absolute magnitudes of the pairs in $g-$band and $r-$band respectively. We divide the sample into 10 percentiles in magnitude $M_r$ and then we compute the median color in each percentile (black dots). Then, we fit a linear function (black line) to the obtained points and select the pairs with larger colors as \textit{Red pairs} and with lower colors as \textit{Blue pairs}. \textit{Upper} and \textit{lower panels} show the total absolute magnitude and the total color distributions of the pairs. The vertical blue line in the absolute magnitude distribution corresponds to the median value of this distribution, which is considered to classify \textit{Higher} and \textit{Lower luminosity} pairs.} 
\label{fig:blue_red}
\end{figure}

The sample of galaxy pairs is obtained from a redshift extended version of the Galaxy Pair Catalog \citep[GPC, ][]{Lambas2012} based on the seventh data release of the Sloan Digital Sky Survey \citep[SDSS-DR7,][]{Abazajian09}. This survey comprises 11,663 square degrees of sky imaged in five wave-bands (\textit{u, g, r, i} and \textit{z}) containing photometric parameters of 357 million objects. The spectroscopic survey is a magnitude limited sample to $r_{\lim} < 17.77$ (Petrosian magnitude) and most of the galaxies span a redshift range $0 < z < 0.25$ with a median redshift of 0.1 \citep{Strauss02}. 

Pair members are selected by requiring $r_p < 25$\,$h^{-1}$\,kpc and $\Delta V < 350$km\,s$^{-1}$ (where $r_p$ is the relative projected separation between the galaxies and $\Delta V$ is the line-of-sight relative velocity). We perform a visual inspection of SDSS images to remove false identifications as described in \citet{Lambas2012}. In order to obtain a larger sample and increase the signal-to-noise ratio for the lensing analysis, we extended the original GPC up to $z=0.2$ applying the same identification criteria.

With the aim of discarding pairs embedded in larger virialized structures as groups and clusters we use the redMaPPer v6.3 members catalog \citep[\href{http://risa.stanford.edu/redmapper/}{redMaPPer catalogs,}][]{Rykoff2014}. This catalog was obtained by the photometric cluster finder algorithm redMaPPer (\textbf{red}-sequence \textbf{Ma}tched-filter \textbf{P}robabilistic \textbf{Per}colation), that identifies galaxy clusters as overdensities of red-sequence galaxies around spectroscopically identified central galaxy candidates. In order to compare GPC pairs with the members catalog we restrict the sample to pairs with $z > 0.08$. The discarded pairs are less than $5\%$. It is worth noting that this criterion only discards galaxy pairs in evolved systems with a defined red-sequence.

Magnitudes are calibrated using $K$-corrections of the publicly available code described in \citet{Blanton2007}. To select pairs comprised by galaxies with similar individual stellar masses we  consider an upper limit for the magnitude difference between the galaxy members, $M_r^2 - M_r^1 < 2.5$, where $M_r^1$ and $M_r^2$ are the absolute magnitudes in $r$-band of the brightest and the second brightest galaxy. The final catalog includes 1911 galaxy pairs in the SDSS-DR7 region ranging from redshift 0.08 to 0.2.

In order to perform the lensing analysis we select the pairs that are included in the CS82 area. The total sample includes 388 pairs. A larger density of pairs is located in the CS82 footprint since more spectroscopic information was available in this field at the time when the catalog was built.  
According to the visual inspection scheme described in \citet{Lambas2012}, pairs are classified considering the interaction between the members into three categories: pairs undergoing merging, $M$; pairs with evident tidal features, $T$; and non disturbed, $N$. In this work we consider $N$ pairs as \textit{Non interacting} and we combine $T$ and $M$ pairs as \textit{Interacting} pairs. This eye-ball classification was performed by one of the authors in order to maintain a unified criteria. The reliability of this procedure was addressed by the comparison with the classification of a subsample of pairs by another author. Then we estimate the classification uncertainty, obtaining $\lesssim 4\%$, according to the differences between the classifications by both authors in the same subsample. The purity of the catalog is therefore guaranteed by the process of visual inspection, in particular for the $T$ and $M$ systems with evidence of physical interaction.

We also classified galaxy pairs into \textit{Red} and \textit{Blue} pairs considering their position in the color-magnitude diagram (Fig. \ref{fig:blue_red}) and into \textit{Higher luminosity} and \textit{Lower luminosity}, according to their total $r-$band luminosity. Pairs with luminosities above the median of the total sample define the \textit{Higher luminosity} subsample and conversely for the \textit{Lower luminosity} one. Redshift, $r_p$ and $\Delta V$ distributions for the samples considered for the lensing analysis are shown in Fig. \ref{fig:pares_dist}. As it can be seen, \textit{Interacting} pairs tend to have lower $r_p$ distances than the \textit{Non interacting} systems. On the other hand $r_p$ distributions for \textit{Red} and \textit{Blue} pairs, as well as \textit{Higher} and \textit{Lower luminosity} pairs, are all consistent with the total sample $r_p$ distribution. The total number of pairs in each subsample is presented in Table \ref{tab:sample} [for the total SDSS-DR7 region and] for the pairs included in the CS82 masked area.

\begin{table}[ht!]
\caption{Number of galaxy pairs in each sample included in the SDSS-DR7 area and in the CS82 masked area.}
\begin{tabular}{@{}ccc@{}}
\hline
\hline
\rule{0pt}{1.05em}%
  Selection criteria   &   SDSS-DR7  & CS82 \\
 \hline
\rule{0pt}{1.05em}%

Total Sample             & 1911 & 388 \\
Non interacting          & 702  & 176\\ 
Interacting pairs        & 1209 & 212\\ 
Red pairs                & 922  & 181\\  
Blue pairs               & 989  & 207\\ 
Higher luminosity pairs  & 1217 & 194 \\
Lower luminosity pairs   & 694  & 194\\ 
\hline         
\end{tabular}
\medskip
\label{tab:sample}
\end{table}

\section{WEAK-LENSING ANALYSIS}
\label{sec:lensing}

\subsection{Shear catalog and stacking technique}

We use the same shear catalog and stacking technique as in \citet{Chalela2018}. The shear catalog is based on the CS82 Survey, which is a joint Canada-France-Brazil project designed with the goal of complementing existing Stripe 82 SDSS $ugriz$ photometry with high quality $i^\prime-$band imaging suitable for weak and strong lensing measurements. This survey is built from 173 MegaCam $i^\prime-$band images and corresponds to an effective area of $129.2$ degrees$^2$, after masking out bright stars and other image artifacts. It has a median Point Spread Function (PSF) of $0.6''$ and a limiting magnitude $i^\prime \sim 24$ \citep{Leauthaud2017}. Shear catalogs were constructed using the same weak lensing pipeline developed
by the CFHTLenS collaboration using the \textit{lens}fit Bayesian shape measurement method \citep{Miller2013}. Further details regarding the lensing catalog can be found e.g. in \citet{Leauthaud2017,Shan2017,Pereira2018,Chalela2018}.

Background galaxies, i.e. galaxies affected by the lensing effect, are selected according to the \citet{Soo2017} photometric redshift catalog. We select galaxies with Z\_BEST > $z_p$+0.1 and ODDS\_BEST $\geq 0.6$, where $z_p$ is the galaxy pair redshift, according to the average redshift of the galaxies, Z\_BEST is the photometric redshift provided by \citet{Soo2017} and ODDS\_BEST expresses the quality of the redshift estimates, from 0 to 1. 

In order to perform the lensing study we apply a stacking analysis to increase the signal. We will summarize this technique in this subsection since it is described in detail in previous works \citep[e.g.,][]{Gonzalez2017,Chalela2017,Chalela2018}. The weak gravitational lensing effect can be characterized by two quantities, the convergence $\kappa$, which accounts for the isotropic stretching of source images, and an anisotropic distortion given by the complex-value lensing shear, $\gamma = \gamma_1 + i \gamma_2$. The tangential component of the shear is related to the surface mass density, $\Sigma(r)$, through \citep{Bartelmann1995}: 
\begin{equation}
\label{Dsigma}
\tilde{\gamma}_{T}(r) \times \Sigma_{\rm crit} = \bar{\Sigma}(<r) - \bar{\Sigma}(r) \equiv \Delta\tilde{\Sigma}(r),
\end{equation}
where we define the density contrast, $\Delta\tilde{\Sigma}$. Here $\tilde{\gamma}_{T}(r)$ is the averaged tangential component of the shear in a ring of radius $r$, $\bar{\Sigma}(<r) $ and $\bar{\Sigma}(r)$ are the average projected mass distribution within a disk and in a ring of radius $r$, respectively and $\Sigma_{\rm crit}$ is defined below (Eq. \ref{eq:sig_crit}). On the other hand, the cross-component of the shear, $\gamma_{\times}$, defined as the component tilted at $\pi$/4 relative to the tangential component, should be zero. We can obtain an unbiased estimate of the shear as: $\langle e \rangle \approx \gamma$, where $\langle e \rangle$ is the averaged ellipticity of background galaxies in annular bins.

The stacking methodology consists in considering an amount of systems to derive the average mass. By combining several lenses, the density of the source galaxies is artificially increased. The density contrast is obtained as the weighted average of the tangential ellipticity of background galaxies of the lens sample:
\begin{equation}
\langle \Delta \tilde{\Sigma}(r) \rangle = \frac{\sum_{j=1}^{N_{Lenses}} \sum_{i=1}^{N_{Sources,j}} \omega_{LS,ij} \times e_{T,ij} \times \Sigma_{\rm{crit},ij}}{\sum_{j=1}^{N_{Lenses}} \sum_{i=1}^{N_{Sources,j}} \omega_{LS,ij}},
\end{equation}
where $\omega_{LS,ij}$ is the inverse variance weight computed according to the weight, $\omega_{ij}$, given by the $lens$fit algorithm for each background galaxy, $\omega_{LS,ij}=\omega_{ij}/\Sigma^2_{\rm crit}$. $N_{Lenses}$ is the number of lensing systems and $N_{Sources,j}$ the number of background galaxies located at a distance $r \pm \delta r$ from the $j$th lens. $\Sigma_{\rm{crit},ij}$ is the critical density for the $i$th source of the $j$th lens, defined as:
\begin{equation} \label{eq:sig_crit}
\Sigma_{\rm{crit},ij} = \dfrac{c^{2}}{4 \pi G} \dfrac{D_{OS,i}}{D_{OL,j} D_{LS,ij}}.
\end{equation}
Here $D_{OL,j}$, $D_{OS,i}$ and $D_{LS,ij}$ are  the angular diameter distances from the observer to the $j$th lens, from the observer to the $i$th source and from the $j$th lens to the $i$th source, respectively. These distances are computed according to the adopted redshift for each galaxy pair (lens) and for each background galaxy (source). $G$ is the gravitational constant and $c$ is the light speed.

Once the density contrast is computed, we take into account a noise bias factor correction as suggested by \citet{Miller2013}, which considers the multiplicative shear calibration factor $m(\nu_{SN},l)$ provided by \textit{lens}fit. For this correction we compute:
\begin{equation}
1+K(z_L)= \frac{\sum_{j=1}^{N_{Lenses}} \sum_{i=1}^{N_{Sources,j}} \omega_{LS,ij} (1+m(\nu_{SN,ij},l_{ij}))}{\sum_{j=1}^{N_{Lens}} \sum_{i=1}^{N_{Sources,j}} \omega_{LS,ij}}
\end{equation}
following \citet{Velander2014,Hudson2015,Shan2017,Leauthaud2017,Pereira2018}. Then, we calibrate the lensing signal as:
\begin{equation}
\langle \Delta \tilde{\Sigma}^{cal}(r) \rangle = \frac{ \langle \Delta \tilde{\Sigma}(r) \rangle }{1+K(z_L)}.
\end{equation}
The projected density contrast profile is computed using non-overlapping concentric logarithmic annuli to preserve the signal-to-noise ratio of the outer region, from $r_{in}=100\,h^{-1}_{70}$\,kpc (where the signal becomes significantly positive) up to $r_{out} = 1.5\,h^{-1}_{70}$\,Mpc. 

Considering results regarding misscentering presented in \citep{Chalela2017} We adopt a luminosity weighted center,  according to the $r-$band galaxy member luminosity. Therefore, the center coordinates is obtained as
\begin{equation}
\begin{split}
R.A._0 = (L_1 \times R.A._1 + L_2 \times R.A._2)/(L_1+L_2), \\
Dec._0 = (L_1 \times Dec._1 + L_2 \times Dec._2)/(L_1+L_2),
\end{split}
\end{equation}
where $R.A.$ and $Dec.$ are the celestial coordinates, $L$ is the $r-$band luminosity, and the sub-indexes $1$ and $2$ refer to the brightest and the second brightest galaxy pair components, respectively. We have tested the center selection by computing the profiles using as centers the brightest galaxy, $R.A._1$, $Dec._1$, and a geometrical center, $R.A._0 = (R.A._1 + R.A._2)\times 0.5 $, $Dec._0 = (Dec._1 + Dec._2)\times 0.5$) and we did not obtain significant differences between the results. This could be accounted for by the adopted inner radius ($r_{in}=100\,h^{-1}_{70}$\,kpc), $\sim 3$ times larger than the galaxy separation, which mitigates any bias introduced by the center selection.

We have considered different logarithm bins, choosing between a total of 7 and 9 radial bins.  We have not observed differences in the density profile parameters, within their uncertainties,  among these choices, therefore we fix the binning in order to obtain the lowest $\chi^2$ value. Given that the uncertainties in the estimated lensing signal are expected to be dominated by shape noise, we do not expect a noticeable covariance between adjacent radial bins and so we treat them as independent in our analyzes. Accordingly, we compute error bars in the profile by bootstrapping the lensing signal using 100 realizations.

We have tested if the Z\_BEST > $z_p$+0.1 selection criterion is sufficient to discard foreground or weak satellite galaxies considering the large uncertainties in the photometric redshifts. Therefore, following \citet{Leauthaud2017} we compute the weak lensing profile for the total sample, with different lens-source separation cuts (Z\_BEST > $z_p$+0.1, Z\_BEST > $z_p$+0.2, Z\_BEST > $z_p$+0.3 and Z\_BEST > $z_p$+0.4). No statistically significant systematic trend is found in this test

\section{Results and discussion}
\label{sec:results}
\renewcommand{\arraystretch}{1.2}
\begin{table*}[ht!]
\caption{Results of the lensing analysis of galaxy pairs.}
\begin{tabular}{@{}c|c|ccc|ccc@{}}
\hline
\hline
\rule{0pt}{1.05em}%
  Selection criteria   &   $N_{Lenses}$  & \multicolumn{3}{c|}{SIS} & \multicolumn{3}{c}{NFW} \\
 &        &$\sigma_{V}$ & $M_{200}$ & $M(r < Me(r_p/2))$ & $R_{200}$ & $M_{200}$ & $M(r < Me(r_p/2))$  \\
 &          & [km\,s$^{-1}$] &  [$10^{12} h^{-1} M_{\odot} $] & [$10^{10} h^{-1} M_{\odot} $] & [$h^{-1}$\,Mpc] & [$10^{10} h^{-1} M_{\odot} $] & [$10^{10} h^{-1} M_{\odot} $]\\

 \hline
\rule{0pt}{1.05em}%

Total Sample            & 388 & $223 \pm 24$ &$6.9 \pm 2.2$  & $16.8 \pm 3.6$ & $0.30 \pm 0.03$ & $7.1 \pm 2.1$  & $2.3 \pm 0.2$ \\ 
Non interacting         & 176 & $200 \pm 38$ &$5.0 \pm 3.0$    & $18.2 \pm 6.9$ & $0.27 \pm 0.05$ & $5.2 \pm 2.8$  & $3.5 \pm 0.8$ \\ 
Interacting pairs       & 212 & $237 \pm 29$ &$8.3 \pm 3.0$  & $13.0 \pm 3.2$ & $0.32 \pm 0.03$ & $8.0 \pm 2.9$  & $1.2 \pm 0.1$ \\ 
Red pairs               & 181 & $264 \pm 28$ &$11.4 \pm 3.7$ & $23.6 \pm 5.0$ & $0.36 \pm 0.03$ & $12.1 \pm 3.6$ & $2.8 \pm 0.3$ \\  
Blue pairs              & 207 & $167 \pm 45$ &$2.9 \pm 2.4$  & $9.6 \pm 5.2$ & $0.22 \pm 0.06$ & $2.7 \pm 2.1$  & $1.6 \pm 0.5$ \\ 
Higher luminosity pairs & 194 & $278 \pm 27$ &$13.2 \pm 3.8$ & $28.2 \pm 5.5$ & $0.36 \pm 0.03$ & $12.7 \pm 3.8$ & $3.3 \pm 0.4$ \\
Lower luminosity pairs  & 194 & $149 \pm 50$ &$2.0 \pm 2.0$  & $6.8 \pm 4.5$ & $0.19 \pm 0.07$ & $1.7 \pm 1.8$  & $1.1 \pm 0.5$ \\ 
\hline         
\end{tabular}
\medskip
\begin{flushleft}
\textbf{Notes.} Columns: (1) Selection criteria; (2) number of galaxy pairs considered in the stack; (3), (4) and (5) results from the SIS profile fit; (6), (7) and (8), results from the NFW profile fit.   
\end{flushleft}
\label{table:results}
\end{table*}

\subsection{Mass profile of stacked galaxy pairs}
We model the derived density contrast profile of galaxy pairs using a Singular Isothermal Esphere (SIS) model and a NFW profile \citep{Navarro1997}. 
The spherically symmetric  NFW profile is derived from numerical simulations, according to the density, averaged over spherical shells, of dark matter halos. This profile depends on two parameters, $R_{200}$, which is the radius that encloses a mean density equal to 200 times the critical density of the Universe, and a dimensionless concentration parameter, $c_{200}$. This density profile is given by:
\begin{equation} \label{eq:nfw}
\rho(r) =  \dfrac{\rho_{\rm crit} \delta_{c}}{(r/r_{s})(1+r/r_{s})^{2}}, 
\end{equation}
where $\rho_{\rm crit}$ is the critical density of the Universe at the average redshift of the lenses, $r_{s}$ is the scale radius, $r_{s} = R_{200}/c_{200}$ and $\delta_{c}$ is the cha\-rac\-te\-ris\-tic overdensity of the halo:
\begin{equation}
\delta_{c} = \frac{200}{3} \dfrac{c_{200}^{3}}{\ln(1+c_{200})-c_{200}/(1+c_{200})}.  
\end{equation}
The mass within $R_{200}$ can be obtained as \mbox{$M_{200}=200\,\rho_{\rm crit} (4/3) \pi\,R_{200}^{3}$}. Lensing formulae for the NFW density profile were taken from \citet{Wright2000}. If both, $R_{200}$ and $c_{200}$, are taken as free parameters to be fitted, we obtain significantly large uncertainties in $c_{200}$ values, due to the lack of information on the mass distribution near the lens center. To overcome this problem, we follow \citet{Uitert2012,Kettula2015} and \citet{Pereira2018}, by using a fixed mass-concentration relation $c_{200}(M_{200},z)$, derived from simulations by \citet{Duffy2008}: 
\begin{equation}
c_{200}=5.71\left(M_{200}/2 \times 10^{12} h^{-1}\right)^{-0.084}(1+z)^{-0.47},
\end{equation}
where we take $z$ as the mean redshift value of the lens sample. The particular choice of this relation does not have a significant impact on the final mass values, which have uncertainties dominated by the noise of the shear profile.

The SIS profile is the simplest density model for describing a relaxed massive sphere with a constant isotropic one dimensional velocity dispersion, $\sigma_V$. At galaxy scales, dynamical studies \citep[eg.,][]{Sofue2001}, as well as strong \citep[eg.,][]{Davis2003} and weak \citep[eg.,][]{Brimioulle2013} lensing observations, are consistent with a mass profile following approximately an isothermal law. The shear, $\gamma(\theta)$, and the convergence, $\kappa(\theta)$, at an angular distance $\theta$ from the lensing system center are directly related to $\sigma_V$ by:
\begin{equation}
\kappa(\theta) = \gamma(\theta) = \dfrac{\theta_{E}}{2 \theta}
\end{equation}
where $\theta_{E}$ is the critical Einstein radius defined as:
\begin{equation}
\theta_{E} = \dfrac{4 \pi \sigma_{V}^{2}}{c^{2}} \frac{D_{LS}}{D_{OS}}.
\end{equation}
Therefore,
\begin{equation}
\tilde{\Sigma}(\theta, \sigma_V) =  \dfrac{\sigma_{V}^{2}}{2 G \theta D_{OL}}. 
\end{equation}
We compute the mass within $R_{200}$, following \citet{Leonard2010}:
\begin{equation}\label{eq:MSIS}
M_{200} =  \dfrac{2 \sigma_{V}^{3} }{\sqrt{50} G H(z)}, 
\end{equation} 
where $H(z)$ is the redshift dependent Hubble parameter. This equation is derived by equating 200$\rho_c$ with the median density computed inside a spherical volume with radius $R_{200}$. 

To derive the parameters of each mass model profile we perform a standard $\chi^{2}$ minimization:
\begin{large}
\begin{equation} \label{eq:chi}
\chi^{2} = \sum^{N}_{i} \dfrac{(\langle \tilde{\Sigma}^{cal}(r_{i})  \rangle - \tilde{\Sigma}(r_{i},p))^{2}}{\sigma^{2}_{\Delta \tilde{\Sigma}}(r_{i})},
\end{equation}
\end{large}\\
where the sum runs over the $N$ radial bins of the profile and the model prediction $p$ refers to either $\sigma_{V}$ for the SIS profile, or $R_{200}$ in the case of the NFW model. Errors in the best-fitting parameters are computed according to the variance of the parameter estimates.

\subsection{Galaxy pairs total masses}

\begin{figure*}
\includegraphics[scale=0.45]{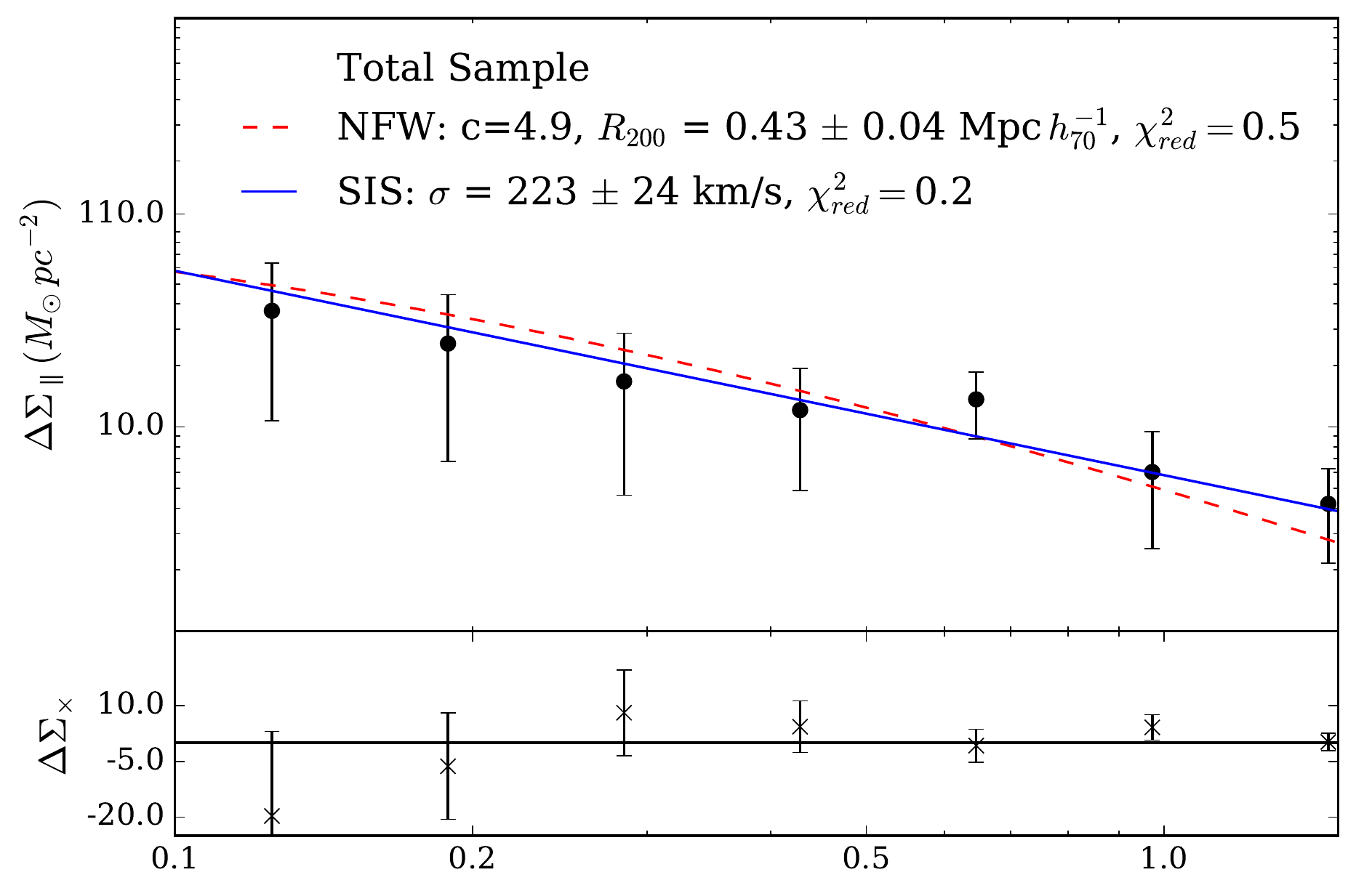}
\includegraphics[scale=0.45]{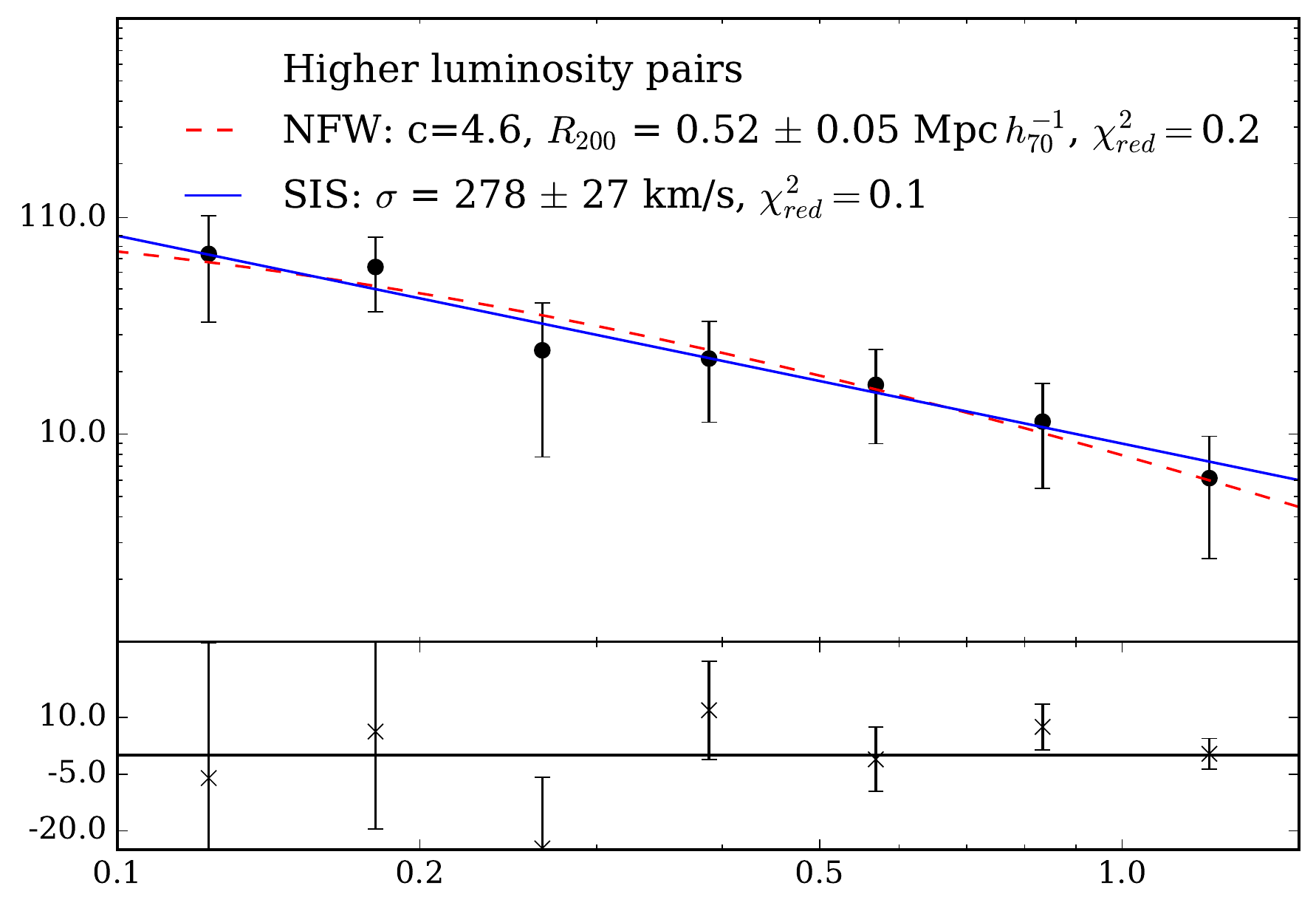}\\
\includegraphics[scale=0.45]{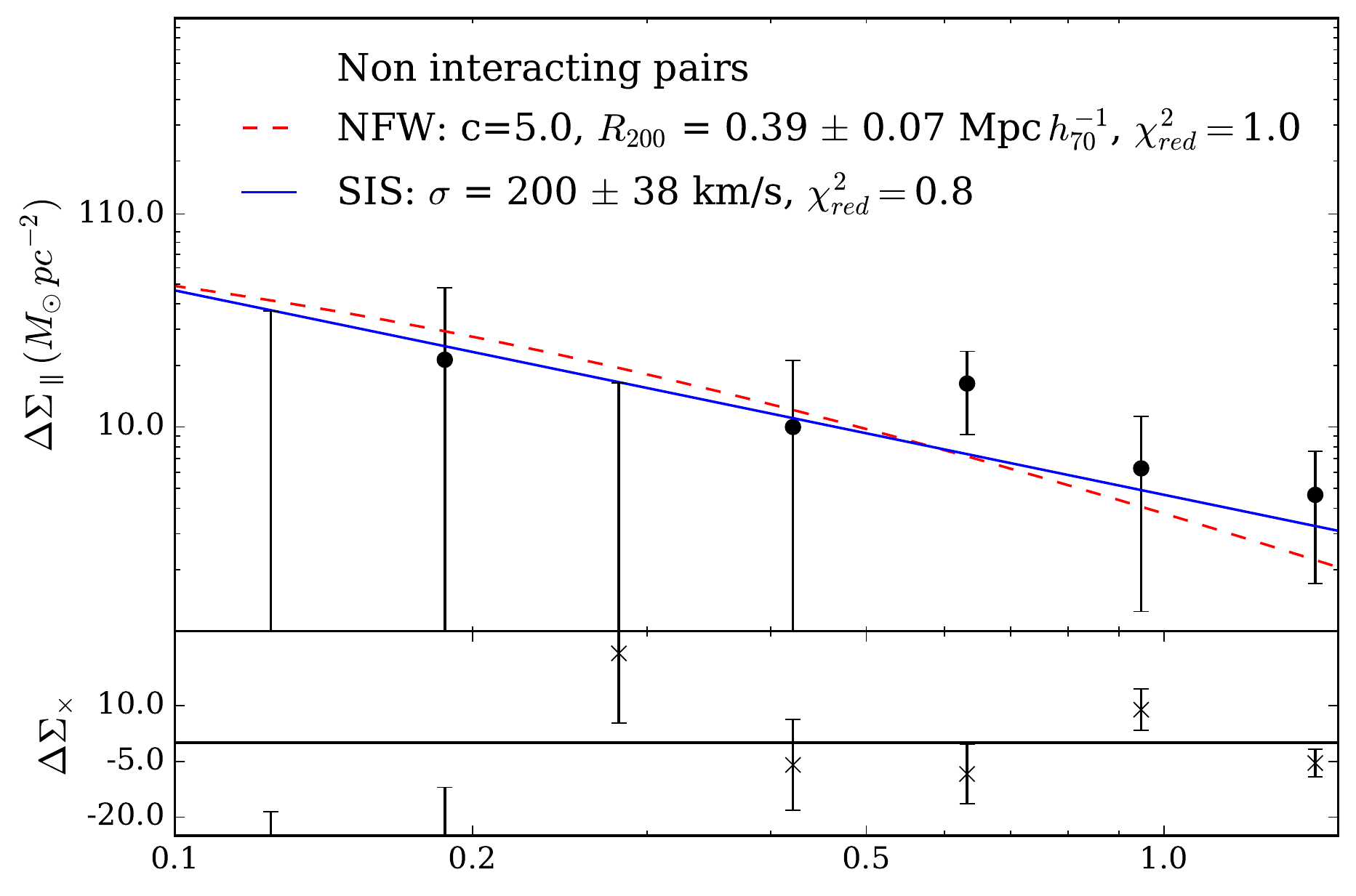}
\includegraphics[scale=0.45]{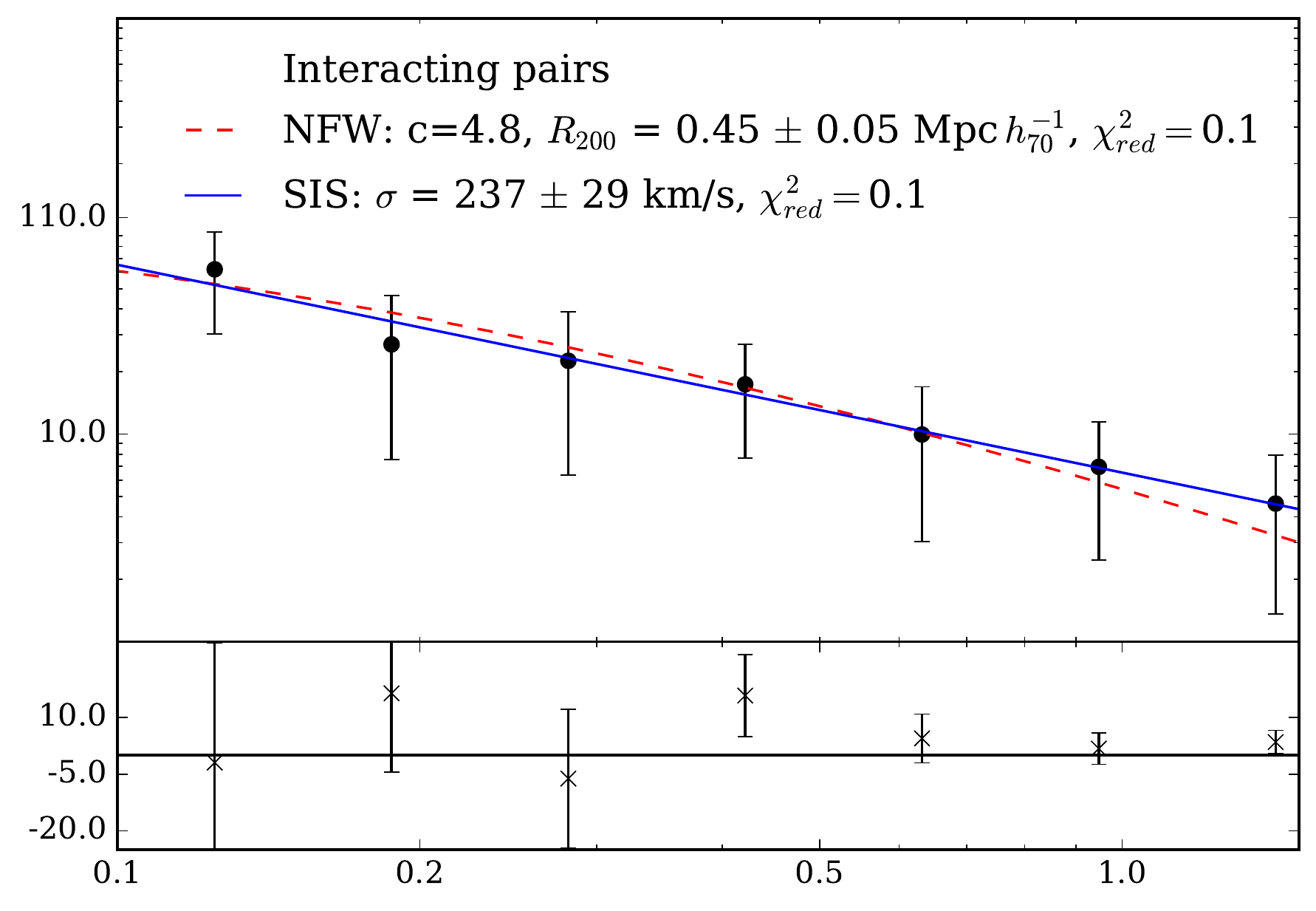}\\
\includegraphics[scale=0.45]{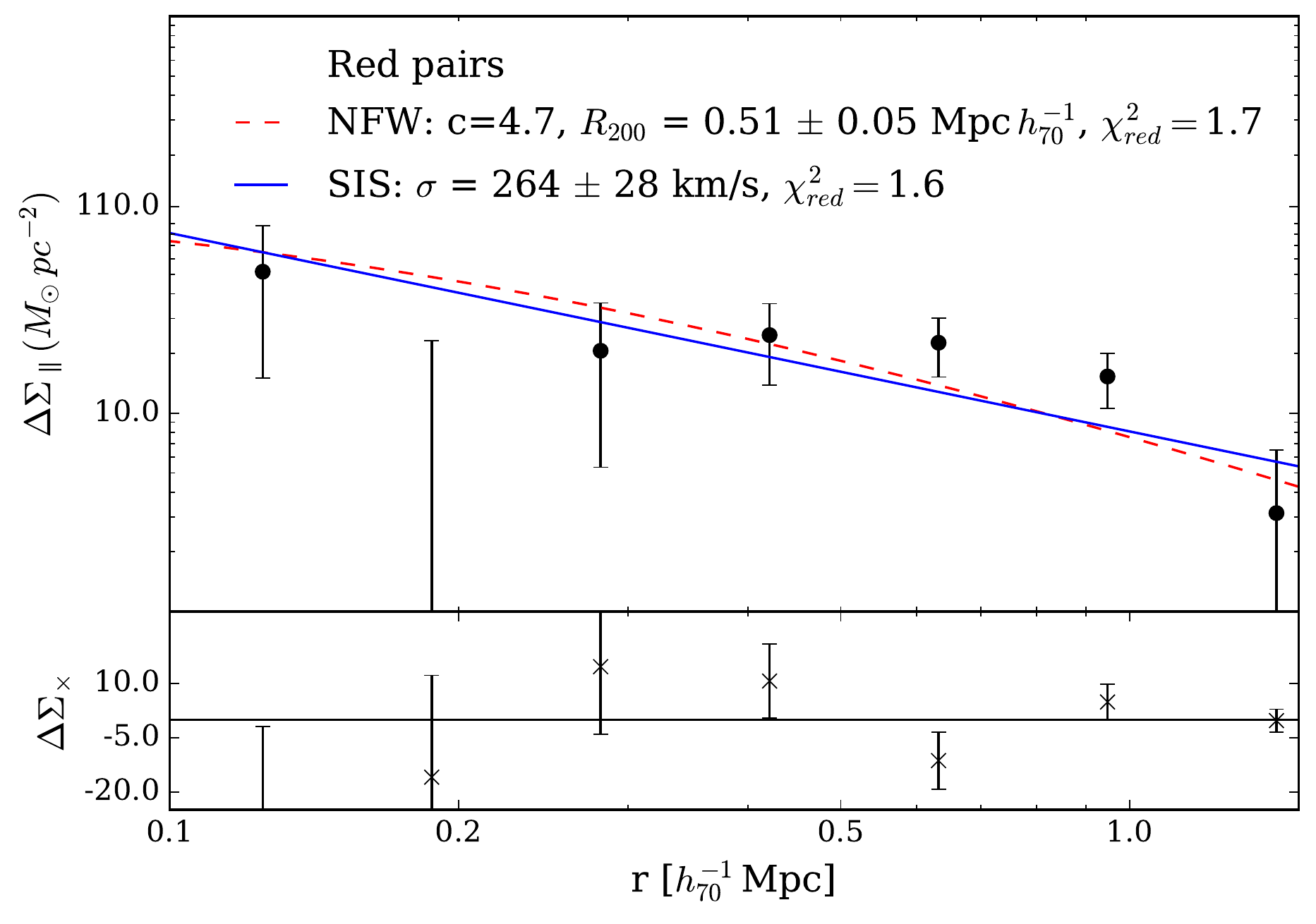}
\includegraphics[scale=0.45]{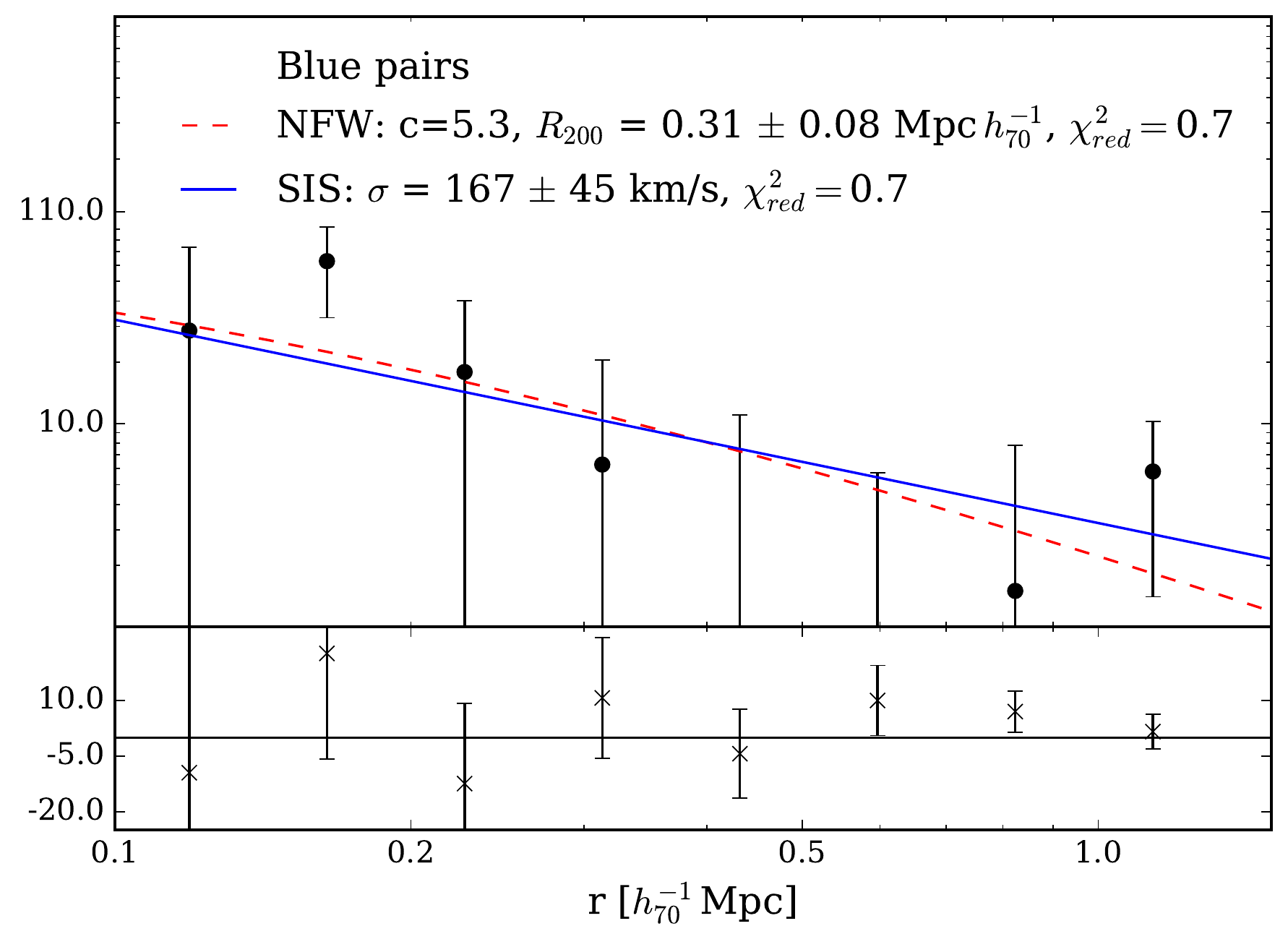}
\caption{Average density contrast profiles obtained for the analyzed galaxy pairs samples. $h_{70}$ corresponds to $h=0.7$.}
\label{fig:profiles}
\end{figure*}

Derived density contrast profiles for each sample are shown in Fig. \ref{fig:profiles} together with the fitted SIS and NFW models. The \textit{Lower luminosity} sample is consistent with no signal ($M_{200}$ masses are poorly determined with $1\sigma$ confidence) and is not shown in the figure, therefore this sample is discarded for the rest of the analysis. In Table \ref{table:results} we show the derived parameters. We could determine, with $\sim 3\sigma$ confidence, masses for the \textit{Total sample} and for \textit{Interacting}, \textit{Red} and \textit{Higher luminosity} pairs. Moreover, there is a significant difference of derived masses between \textit{Higher} and \textit{Lower luminosity} pairs, since galaxy pairs with larger luminosity are more massive systems. \textit{Non interacting} galaxy pairs tend to have lower total luminosities than \textit{Interacting} pairs which is in agreement with the lower lensing mass determined for \textit{Non interacting} pairs. On the other hand, $Red$ pairs are more evolved systems since this sample includes more elliptical members. Hence, it is expected that these systems have larger masses than $Blue$ pairs, as is observed from the lensing analysis. 

\begin{figure*}
\begin{center}
\includegraphics[scale=0.47]{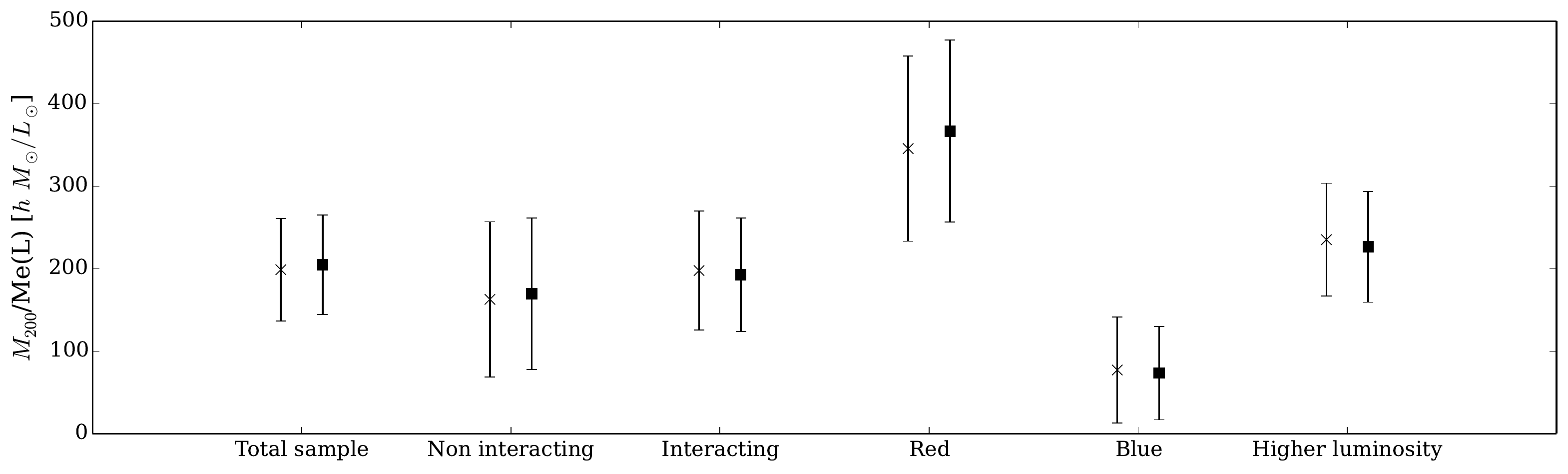}
\caption{Mass to light ratios for the analyzed samples. Masses are the $M_{200}$ derived according to the lensing analysis, considering SIS (crosses) and NFW (squares) models. Luminosity is computed according to $r-$band absolute magnitudes, adopting the median of the luminosity distribution for each sample. }
\label{fig:ML}
\end{center}
\end{figure*}

According to the derived masses we compute the mass-to-light ratio for the analyzed samples (Fig. \ref{fig:ML}). We consider the median luminosity according to the $r-$band absolute magnitude, Me(L). We obtain for the \textit{Total sample} a $M_{200}$/Me(L) $\sim 200 h\,M_\odot/L_\odot$. This value is in agreement with mass-to-light ratio determinations for groups of similar masses as the analyzed pairs \citep{Proctor2011,Girardi2002}. Derived $M_{200}$/Me(L) values for \textit{Interacting}, \textit{Non-interacting} and \textit{Higher luminosity} pairs are in agreement with the derived value for the \textit{Total sample}. On the other hand \textit{Red} pairs show a larger $M_{200}$/Me(L), since this sample includes a larger fraction of elliptical members. It can be noticed from Fig. \ref{fig:ML} that both lensing mass estimates, SIS and NFW, are in excellent agreement, indicating that the available data do not allow us to distinguish between the fitted models. 

\subsection{Comparison with dynamical estimates}

\begin{figure}
\begin{center}
\includegraphics[scale=0.5]{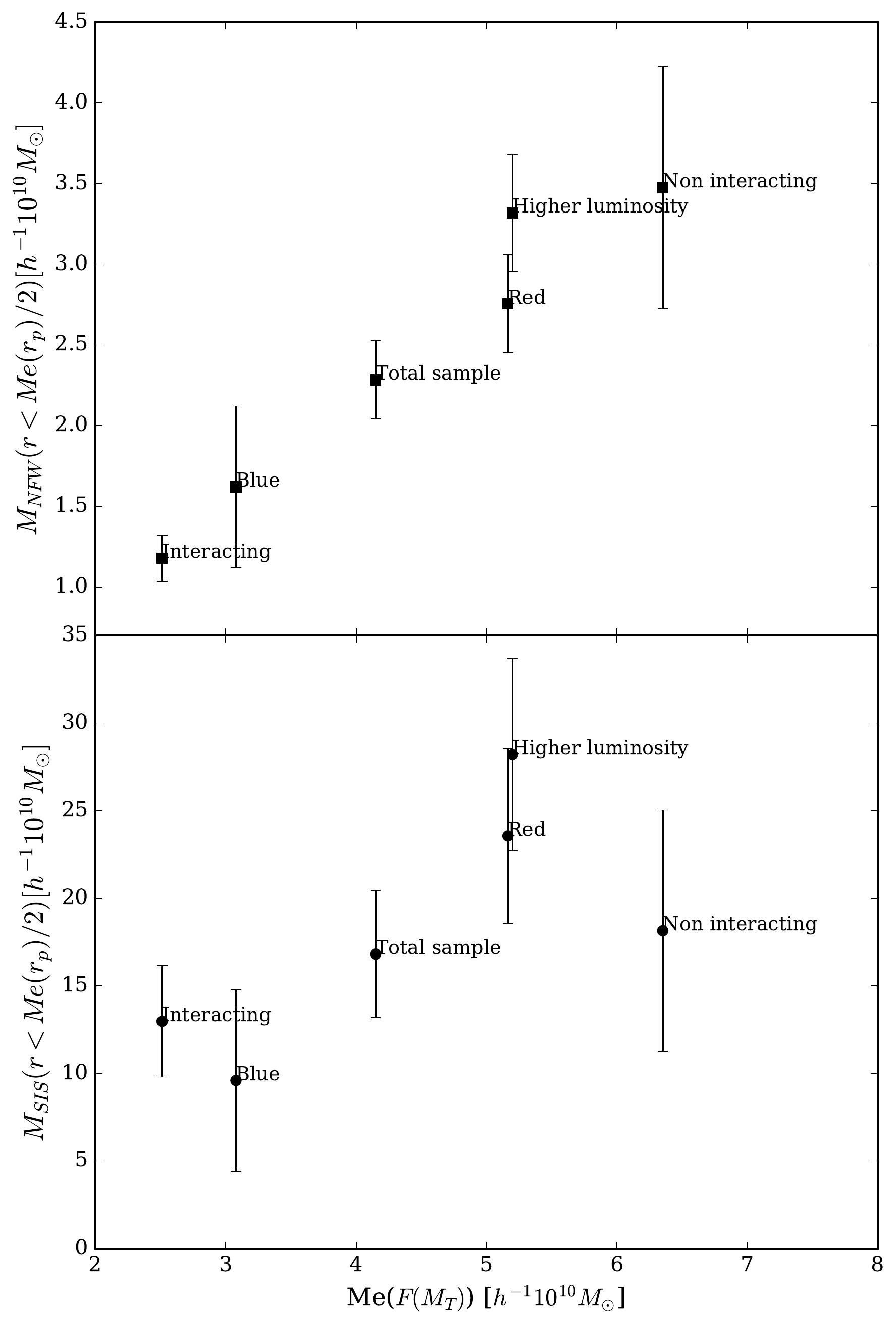}
\caption{Derived lensing masses compared to the median $F(M_T)$ defined according to Eq. \ref{eq:dyn}, for the different galaxy pair samples.}
\label{fig:Mvir}
\end{center}
\end{figure}

In order to compare our mass determinations to dynamical estimates, we analyze the relation between our lensing results and an \textit{indicative mass}, defined as \citep{Faber1979}:
\begin{equation} \label{eq:dyn}
F(M_T) = \frac{r_p \Delta V^2}{G}, 
\end{equation}
which is related to the dynamical mass of a pair of objects in relative Keplerian motion. This parameter has been used to estimate the total mass enclosed within $r_p$, although several assumptions regarding galaxy orbits are required \citep{Nottale2018}. Also, this approach considers that the pair is isolated and lacks dynamical friction effects. This is not entirely accurate since the large scale structure has significant effects on the dynamics of galaxy pairs \citep{Moreno2013,Mesa2018} and variations of the internal structure of the galaxies can modify their center of mass motion.

We derive the masses within a radius Me($r_p$)/2 (where Me($r_p$) is the median value of the projected distances derived for each sample) using the lensing results. The SIS mass inside a three dimensional radius Me($r_p$)/2 is \citep{Wright2000}:
\begin{equation}
M_{SIS}(r<Me(r_p)/2) = \frac{\sigma_V^2 Me(r_p)}{G}.
\end{equation}
The NFW mass is obtained by:
\begin{equation}
M_{NFW}(r<Me(r_p)/2) = 4 \pi \int^{Me(r_p)/2}_0 \rho(r) r^2 dr,
\end{equation}
where $\rho(r)$ is computed according to Eq. \ref{eq:nfw}. The results are shown in Table \ref{table:results}. As expected, given that the SIS profile is steeper than the NFW profile, SIS masses within Me($r_p$)/2 are significantly larger than the NFW determinations.

In Fig. \ref{fig:Mvir} we show the relation between $M_{SIS}(r<Me(r_p)/2)$ and $M_{NFW}(r<Me(r_p)/2)$ masses and the median values of $F(M_T)$ for each sample. 

It is worth noting that the SIS and NFW mass profile parameters are obtained from the lensing signal at significantly larger radii compared to $r_p$. Therefore, the masses extrapolated down to $r_p$ will be highly dependent on the assumed profile, as can be seen in the figure. In any case, there is a clear correlation between the indicative mass and the lensing masses at $r_p$.
We find that SIS masses are $\sim 5$ times the median $F(M_T)$ values while NFW masses are roughly half of $F(M_T)$. The indicative dynamical mass is expected to set a lower limit for the dynamical mass inside $r_p$. Therefore, taken at face value, these results would favor the SIS profile compared to the NFW one, or could point to another scaling at small radii. In any case, astrophysical processes in interacting pairs, as well as the inclusion of gravitationally unbound pairs in the sample, can significantly bias dynamical estimates. Further analysis including virial mass determinations considering the probability distribution for the projected $r_p \Delta V^2$ value, and a comparison with other density distribution models, would be important to deepen our understanding of the mass distribution in close galaxy pairs. 

\subsection{Comparison with Yang groups}

\begin{figure}
\begin{center}
\includegraphics[scale=0.5]{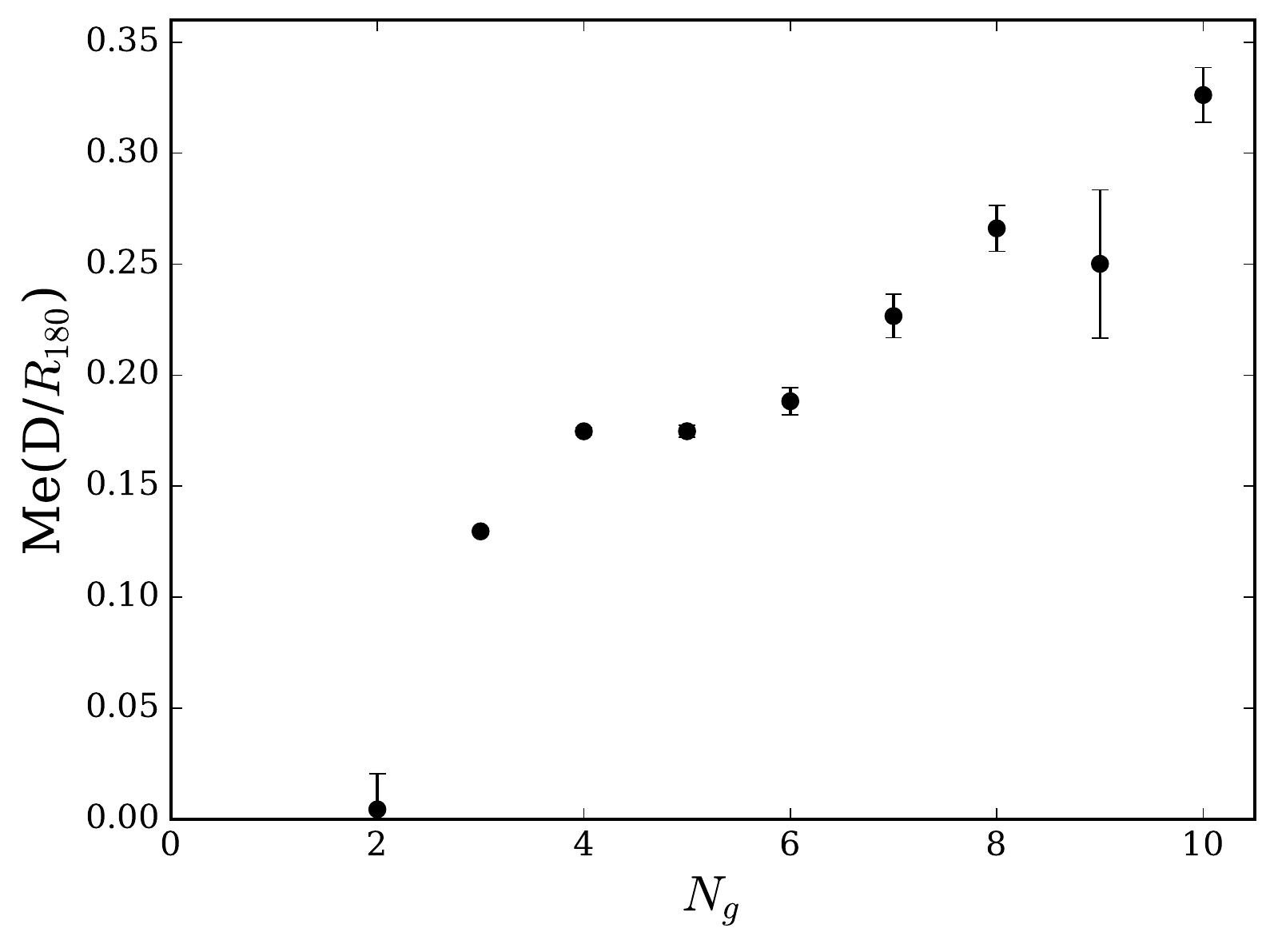}\\
\caption{Median distances of the pairs scaled with the radius, $R_{180}$, to the group centers, according to the group number of members identified, $N_g$. Error bars are computed according to the root-mean-square deviation. For these systems median $R_{180}$ values range from 400\,kpc to 1\,Mpc.}
\label{fig:dist}
\end{center}
\end{figure}

\begin{figure*}
\begin{center}
\includegraphics[scale=0.5]{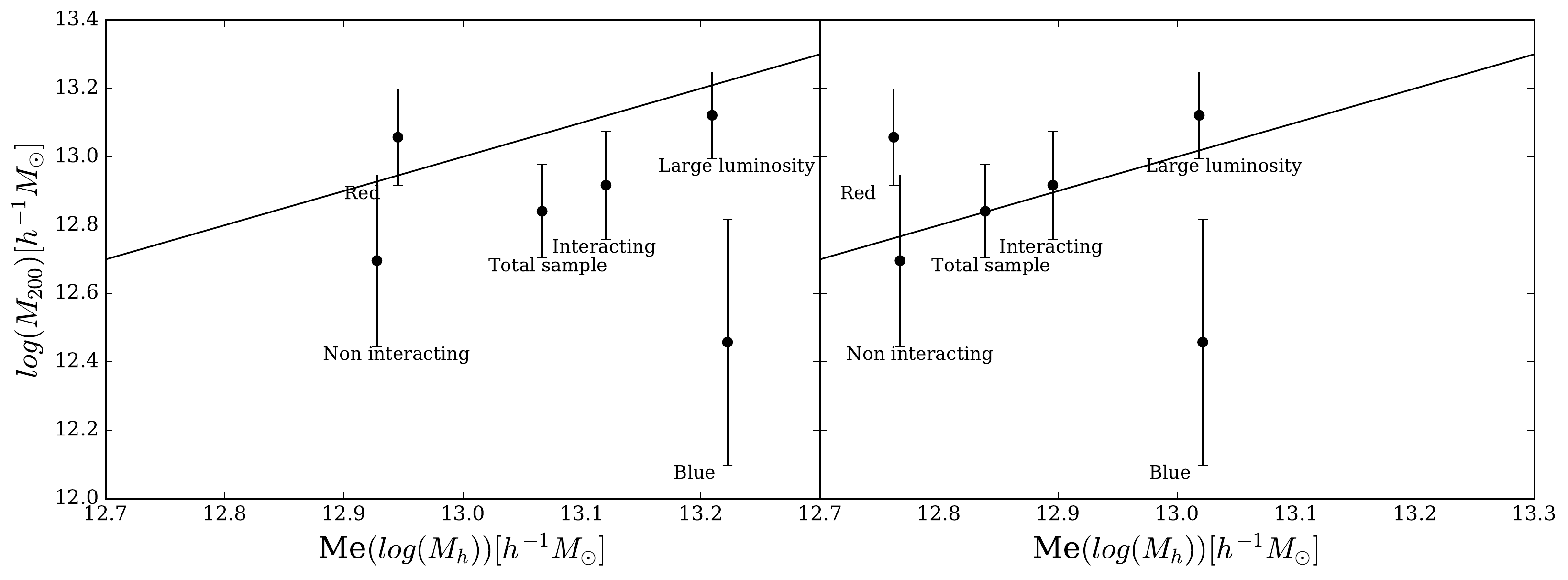}
\caption{Derived SIS lensing masses ($M_{200}$) compared to the median masses derived by Yang, for the correlated pairs (\textit{left panel}) and restricting the sample to pairs in groups with only two members (\textit{right panel}). The solid black line corresponds to the identity function.}
\label{fig:yang}
\end{center}
\end{figure*}

In this section we compare derived lensing masses with those available in a galaxy group catalog, in order to validate our results with other independent determinations. We chose to compare our mass estimates with group masses determined by \citet{Yang2012}. We select this catalog mainly because it has reliable groups with low number of members and given our experience with these galaxy systems \citep{Rodriguez2015,Gonzalez2017}. This group catalog was constructed using the adaptive halo-based group finder presented in \citet{Yang2005}, updated to SDSS DR7. The idea of this algorithm is to assign galaxies into groups according to their common dark matter halo. We refer the reader to \citet{Yang2005} for a complete description of the method and to \citet{Campbell2015} for a brief description and a deep analysis of the resulting groups identified by this algorithm.

For our comparison, it is important to take into account that the group mass is assigned according to a luminosity ranking and using an iterative relation between mass and luminosity. For this assignment, this method requires knowledge of the halo mass function and assumes the existence a one-to-one relation between mass and luminosity for a given comoving volume and a given halo mass function. Group masses have a scatter of approx 0.3 dex \citep{Yang2007} and, as pointed out by \citet{Campbell2015}, the applied method is fundamentally limited in two ways: (i) the intrinsic scatter in the relation between group luminosity and halo mass; (ii) any errors in group membership determination will generally result in errors in group luminosity, and thus in the inferred halo mass. In particular, regarding to membership allocation errors, two failure modes are associated with this step, fragmentation and merging. Fragmentation refers to galaxies of the same group that are fractured in two or more groups, while merging refers to galaxies of separate groups that are assigned to one group. 

In order to carry out the comparison we use galaxy pairs, covering the total SDSS-DR7 imaging area, described in Sec. \ref{sec:data}. The sample includes 1911 pairs, of which 1656 have at least one galaxy that is identified as a member of a group. 1177 have both galaxies identified in the same group, $95 \%$ of the this sample are in systems with less than 10 members, while $\sim 60 \%$ (706 pairs) are identified in groups with only 2 members. This may be due to errors regarding membership allocation to the identified groups and/or because GPC includes galaxy pairs that are not isolated and are members of larger systems. For lower mass systems, merging is more important in the membership assignment than fragmentation. This will result in an overestimate of its group mass. On the other hand, non isolated pairs located near the center of the groups will bias the lensing signal to larger values and therefore, larger mass estimates. In Fig. \ref{fig:dist} we show the median distances of the pair center to the group center according to the number of members. As it can be noticed pairs are at a median distance from the group center lower than $0.35 \times R_{180}$, where $R_{180}$ are computed according the halo group masses and is the radius that encloses a mean density equal to 180 times the critical density of the Universe. Hence, pairs are mainly located near the group centers.

In Fig. \ref{fig:yang} we show the lensing SIS masses compared to median group masses, $M_g$, computed by \citet{Yang2012} for each sample. In the left panel we show the comparison considering the groups that hold the two galaxies of the pair and have less than 10 identified members. Median group masses are biased to larger values ($\langle M_g - M_{200} \rangle = 4\times10^{12} h^{-1} M_{\odot} $) indicating that merging in the membership assignment could play an important role. In the right panel we compare the masses considering only groups that include both galaxies and with only two members, a good agreement is obtained for this relation ($\langle M_g - M_{200} \rangle = -0.44\times10^{12} h^{-1} M_{\odot} $). Although that groups with low and intermediate masses have large uncertainties in their mass determinations, the median value for the total sample is in excellent agreement with the lensing estimate. For this sample, larger differences are observed for \textit{Red} and \textit{Blue} pairs, which could be due to the intrinsic scatter in the relation between group luminosity and halo mass. Since group masses are determined considering a non-color dependent mass-to-light ratio, this could bias mass determinations. As it was shown in Fig. \ref{fig:ML}, \textit{Red} and \textit{Blue} pairs show differences with the derived $M_\odot/L_\odot$ compared to the other samples, therefore halo mass determinations for individual groups with a low number of galaxy members might be biased according to their color.

\section{Summary and conclusions}
\label{sec:conc}

In this work we present a lensing analysis of galaxy pairs using weak lensing techniques. We determined lensing masses for different samples selected according to pair luminosity, color and interaction visual classification. The derived mass-to-light ratios for the different samples are in agreement with $\sim 200\,h\,M_\odot/L_\odot$, except for the color selected samples, where the \textit{Red} pair sample shows a larger mass-to-light ratio than \textit{Blue} pairs. This is in agreement with other works that report a correlation between dynamical mass-to-light ratios and colors for individual galaxies \citep{Zandivarez2011,Sande2015}, consistent with the fact that a galaxy $M/L$ ratio strongly depends on age, metallicity, and stellar initial mass function. Furthermore, according to \citet{Wang2014}, galaxy pairs composed by two early-type galaxies reside in high mass halos and are often associated to more massive filaments \citep{Mesa2018}. 

Derived lensing masses of \textit{Interacting} pairs are also significantly larger than those masses of \textit{Non-interacting} pairs. The inclusion of line-of-sight pairs in the latter could bias mass estimates to lower values, since truly bound galaxy systems tend to reside in more massive halos than isolated galaxies. 

Lensing masses obtained within the median projected distance of the galaxies for each sample, were compared to an indicative of the dynamical mass, $F(M_T)$. Due to projection effects, this parameter provides a lower limit of the dynamical mass enclosed by the galaxy pair. There is a clear correlation between $F(M_T)$ and lensing derived NFW and SIS model masses. However, assuming an NFW profile, lensing masses are $\sim 50\%$ lower than $F(M_T)$, while SIS masses exceed $F(M_T)$ values by a factor $\sim 5$. Astrophysical processes undergoing in interacting systems could bias low the projected relative velocity, $\Delta V$. This could lead to underestimates of the dynamical mass, and in particular, dynamical friction will drives orbital angular momentum transfer into the internal regions of halos \citep{Gabbasov2014}. On the other hand, gravitationally unbound pairs would result in larger $\Delta V$ values. Also, it is important to take into account that the derived lensing parameters are obtained from the projected density distribution of the halo at considerably larger radii compared to the projected distance between the galaxies. Therefore, significant deviations of the density distribution from single NFW and SIS profiles could be present.

We also compare lensing masses with \citet{Yang2012} group determinations. We notice that wrongly merged groups resulting in erroneous membership allocation might strongly bias high the group masses. Nevertheless, when the group sample is restricted to galaxy systems with only two members, a very good agreement between the lensing estimates and group masses is obtained. Although, \textit{Red} and \textit{Blue} pairs show large differences between group and lensing masses, we remark that this is likely to be caused by a single mass-to-light ratio adopted by \citet{Yang2012}, regardless of galaxy color.

Galaxy pairs are the simplest systems that can provide valuable information on galaxy morphology transformations. They are also important in galaxy group and cluster formation since pairs are formed at the center of more massive dark matter halos than single galaxies \citep{Wang2014}. In this work we derive lensing masses for a sample of close pairs and we find that \textit{Red} pairs show larger masses and a higher mass-to-light ratio. In a future work we plan to constrain the derived results with hydrodynamical simulations in order to explore galaxy formation and evolution scenarios. The study of these galaxy systems will contribute to a better understanding of galaxy transformation and group formation.

\begin{acknowledgements}

The authors thank the anonymous referee for its useful comments that helped improved this paper. 

This work was partially supported by the Consejo Nacional
de  Investigaciones  Cient\'ificas  y  T\'ecnicas  (CONICET, Argentina), the Secretar\'ia de Ciencia y Tecnolog\'ia de la Universidad Nacional de C\'ordoba (SeCyT-UNC, Argentina), the Secretar\'{\i}a de Ciencia y T\'ecnica de la Universidad Nacional de San Juan, the Brazilian Council for Scientific and Technological Development (CNPq) and the Rio de Janeiro Research Foundation (FAPERJ). 
We acknowledge the PCI BEV fellowship program from
MCTI and CBPF.
MM acknowledges the hospitality of Fermilab. 
Otag Icram. Fora Temer.  \\

This  work  is  based  on  observations  obtained  with
MegaPrime/MegaCam,  a  joint  project  of  CFHT  and
CEA/DAPNIA, at the Canada–France–Hawaii Telescope
(CFHT), which is operated by the National Research
Council (NRC) of Canada, the Institut National des Sciences de l’Univers of the Centre National de la Recherche Scientifique (CNRS) of France, and the University of
Hawaii. The Brazilian partnership on CFHT is managed
by the Laborat\'orio Nacional de Astrof \'isica (LNA). We
thank the support of the Laborat\'orio Interinstitucional de
e-Astronomia (LIneA). We thank the CFHTLenS team
for their pipeline development and verification upon which
much of the CS82 survey pipeline was built.\\

We also used SDSS data. Funding for the SDSS and SDSS-II has been provided by the Alfred P. Sloan Foundation, the Participating Institutions, the National Science Foundation, the U.S. Department of Energy, the National Aeronautics and Space Administration, the Japanese Monbukagakusho, the Max Planck Society, and the Higher Education Funding Council for England. The SDSS Web Site is http://www.sdss.org/.
The SDSS is managed by the Astrophysical Research Consortium for the Participating Institutions. The Participating Institutions are the American Museum of Natural History, Astrophysical Institute Potsdam, University of Basel, University of Cambridge, Case Western Reserve University, University of Chicago, Drexel University, Fermilab, the Institute for Advanced Study, the Japan Participation Group, Johns Hopkins University, the Joint Institute for Nuclear Astrophysics, the Kavli Institute for Particle Astrophysics and Cosmology, the Korean Scientist Group, the Chinese Academy of Sciences (LAMOST), Los Alamos National Laboratory, the Max-Planck-Institute for Astronomy (MPIA), the Max-Planck-Institute for Astrophysics (MPA), New Mexico State University, Ohio State University, University of Pittsburgh, University of Portsmouth, Princeton University, the United States Naval Observatory, and the University of Washington.
\end{acknowledgements}

\bibliography{references}

\end{document}